\documentclass[prl,twocolumn,showpacs,amsmath,amssymb]{revtex4-1}
\usepackage{color}
\usepackage{ulem}
\usepackage{graphicx}
\usepackage{epstopdf}
\usepackage{bm}
\usepackage{amssymb}
\usepackage{amsmath}
\usepackage[colorlinks,citecolor=blue]{hyperref}
\usepackage{epsfig}

\begin{document}

\title{Photocurrents in gyrotropic Weyl semimetals}

\author{L.\,E.\,Golub$^+$, E.\,L.\,Ivchenko$^+$, B.\,Z.\,Spivak*}

\address{$^+$Ioffe Institute, 194021 St.\,Petersburg, Russia\\
*Department of Physics, University of Washington, Seattle, WA 98195, USA }

\begin{abstract}{
We present results of a theoretical study of photocurrents in the Weyl semimetals belonging to the gyrotropic symmetry classes. We show that, in weakly gyrotropic symmetry classes C$_{nv}$ ($n = 3,4,6$), the circular photocurrent transverse to the incidence direction appears only with account, in the electron effective Hamiltonian, for both linear and quadratic or cubic in quasi-momentum spin-dependent terms as well as a spin-independent term resulting in the tilt of the cone dispersion. A 
polarization-independent magneto-induced photocurrent is predicted which is also allowed in gyrotropic systems only. For crystals of the C$_{2v}$ symmetry, we consider a microscopic mechanism of the photocurrent in a quantized magnetic field which is generated under direct optical transitions between the ground and the first excited magnetic subbands. It is shown that this current becomes nonzero with allowance for anisotropic tilt of the dispersion cones.}
\end{abstract}

\maketitle

\section{Introduction}
A distinctive property of the symmetry of a gyrotropic crystal is the presence of components of the polar (${\bm R}$) and axial (${\bm L}$) vectors which transform according to equivalent representations of the crystal point group. This property allows for effects that connect the physical quantities $R_{\alpha}$ and $L_{\beta}$ by a second-rank pseudo-tensor  
$C_{\alpha \beta}$ or relate an invariant physical quantity $I$ with the products $R_{\alpha} L_{\beta}$. The most famous of such effects is the natural optical activity, or the rotation of the polarization plane of the light wave propagating in a gyrotropic medium. It is described by linear terms in the expansion of the permittivity tensor $\varepsilon_{\lambda \nu} (\omega, {\bm q})$ in powers of the light wave vector ${\bm q}$: a gyrotropic correction to the vector of electric displacement field $\delta {\bm D}$ can be represented as a vector product ${\rm i} \left({\bm E} \times {\bm g} \right)$, where ${\bm E}$ is the electric field of the light wave and ${\bm g}$ is the gyration vector linearly related to the vector ${\bm q}$ \cite{LanLif}. Another gyrotropic effect is the appearance of an electric photocurrent ${\bm j}$ proportional to the vector product ${\rm i} ({\bm E} \times {\bm E}^*)$. It was called the circular photogalvanic effect (CPGE), predicted independently in Refs.~\cite{IvchPik,Belin} and observed for the first time on a bulk tellurium crystal in Ref.~\cite{Asnin} and on a quantum-well structure GaAs/AlGaAs in \cite{Ganichev,Ganichev2}.

In the theoretical works \cite{Moore,Patrick,Koenig} the CPGE is studied in the Weyl semimetals. It is established that the contribution of each Weyl node to the circular photocurrent takes the universal form \cite{Moore}
\begin{equation} \label{Gamma00}
{\bm j} = {\cal C} \Gamma_0\tau_p {\rm i}\left( {\bm E} \times {\bm E}^* \right)\:,
\end{equation}
where $\Gamma_0 = \pi e^3 /3 h^2$, $e$ is the electron charge, $h$ is the Planck constant, ${\cal C} = \pm 1$ is the chirality (or topological charge) of the node, $\tau_p$ is the electron momentum relaxation time. The universality of Eq.~(\ref{Gamma00}) means that, with the exception of the factor $\tau_p$, the right-hand side of this equation contains a numerical factor $\pi/3$ and the world constants $e$ and $h$. Ching-Kit Chan et al. \cite{Patrick} have considered a pair of Weyl nodes with opposite chiralities. They have shown that the contributions to the circular photocurrent do not cancel each other provided that, in addition to the terms $A_{\alpha \beta} \sigma_{\alpha} k_{\beta}$,  the effective electron Hamiltonian contains the tilt term  ${\bm a}\cdot {\bm k}$ with the vector ${\bm a}$ different in the different nodes ($\sigma_{\alpha}$ are the Pauli spin matrices, ${\bm k}$ is the electron wave vector referred to the node ${\bm k}_W$). However, in this case the equation for the photocurrent loses its universality. Recently Qiong Ma et al. \cite{ExpLee} have observed a circular photocurrent in the TaAs crystal under excitation by CO$_2$ laser radiation.

In the present work, we analyze how the presence of a reflection plane in the point-symmetry group of a gyrotropic crystal affects the CPGE and discuss the influence of a magnetic field on the photocurrents in Weyl gyrotropic semimetals. A particular attention is paid to the crystal classes 
C$_{4v}$ and C$_{2v}$, since it has been established that, among the Weyl semimetals, monopniktides TaAs, NbP, NbAs \cite{TaAs,NbP,NbAs} and tungsten telluride WTe$_2$ \cite{WTe2a,WTe2} have these symmetries.
\section{Gyrotropic crystals in the absence of a magnetic field}
Let us consider a crystal with the effective electron Hamiltonian near the point ${\bm k}_W$ given  in the form
\begin{equation} \label{Hamilt}
{\cal H} = {\bm d}({\bm k})\cdot {\bm \sigma} + d_0({\bm k}) \sigma_0\:,
 \end{equation}
where $\sigma_0$ is the identity matrix of dimension 2, $d_l({\bm k})$ $(l=0,x,y,z)$ are functions whose expansions in powers of ${\bm k}$ contain no terms of the zero order. The eigenenergies of this Hamiltonian take on the values $E_{\pm, {\bm k}} = d_0({\bm k}) \pm d({\bm k})$, where $d(\bm k) = |{\bm d}(\bm k)|. $ Hereafter the energy is referred to the electron energy at the Weyl point. Usually, for simplicity only linear terms are taken into account in $d_0({\bm k})$: $d_0({\bm k}) = {\bm a} \cdot {\bm k}$, where ${\bm a}$ is some vector. This term describes the tilt of the Weyl cone. 

Under direct optical transitions in the vicinity of the ${\bm k}_W$ point, the following photocurrent is generated 
\begin{equation} \label{general}
{\bm j} = e \sum\limits_{\bm k} \tau_p \frac{2}{\hbar} \frac{\partial d({\bm k})}{\partial {\bm k}} W_{+-}({\bm k})\:,
\end{equation}
where the rate of optical transitions per unit volume per unit time is given by
 \begin{equation} \label{W+-}
W_{+-} = \frac{2 \pi}{\hbar} \left\vert M_{+-} \right\vert^2 F({\bm k})  \delta \left( 2d - \hbar \omega \right)\:,
\end{equation}
$M_{+-}$ is the matrix element of the optical transition that does not depend on $d_0({\bm k})$, 
\begin{equation} \label{tilt}
F({\bm k}) =  f \left(E_{-, {\bm k}}\right) - f \left(E_{+, {\bm k}} \right)\:,
\end{equation}
$f \left( E \right)$ is the equilibrium Fermi-Dirac distribution function. The factor 2 in Eq.~(\ref{general}) takes into account the contributions to the current from the photoelectrons and photoholes. Similarly to Ref.~\cite{Moore}, one can show that, under the circularly polarized excitation, the 
polarization-dependent contribution to the square modulus of the matrix element is proportional to the Berry curvature
\begin{equation}
\left\vert  M_{+-} \right\vert^2_{\rm circ} = \frac{2 e^2d^2}{(\hbar\omega)^2} |{\bm E}|^2 {\bm \varkappa}\cdot{\bm \Omega} \:,
\end{equation}
where ${\bm \varkappa} = {\rm i} ({\bm e} \times {\bm e}^*)$, ${\bm e}$ is a unit polarization vector, the Berry curvature is related to the vector ${\bm d}({\bm k})$ by
\begin{equation}
\Omega_{j} = \frac{{\bm d}}{2 d^3} \cdot \left(  \frac{\partial {\bm d}}{\partial k_{j+1}} \times \frac{\partial {\bm d}}{\partial k_{j+2}} \right) \:,
 \end{equation}
and the cyclic permutation of the indices is assumed. We note that, taking into account the energy conservation law, the argument $E_{\pm, {\bm k}}$ of the distribution function can be replaced by
$d_0({\bm k}) \pm \hbar \omega/2$. Therefore, for a fixed frequency $\omega$, the difference of occupation numbers (\ref{tilt}) is a function of the scalar $d_0({\bm k})$.

We introduce the pseudotensor ${\bm \gamma}$ which describes the CPGE in accordance with
\begin{equation}
j_{\alpha} = \gamma_{\alpha \beta} \varkappa_{\beta} |{\bm E}|^2\:.
\end{equation}
The structure of the second-rank pseudotensor ${\bm \gamma}$ for all 18 gyrotropic classes is known, see e.g. \cite{Sirotin}. The purpose of our analysis is to find out the simplest form of the Hamiltonian (\ref{Hamilt}) which satisfies the two requirements: (i) it leads to a nonzero contribution of the node ${\bm k}_W$ to the $\gamma_{\alpha \beta}$ component which is allowed by the crystal symmetry group $F$, and (ii) this contribution does not disappear after the summation over the star of the vector ${\bm k}_W$.

First we consider {\it gyrotropic classes that do not contain reflection planes}, and take into account in the Hamiltonian (\ref{Hamilt}) only terms linear in ${\bm k}$: $d_{\alpha} = A_{\alpha \beta} k_{\beta}$, with the chirality of the node ${\bm k}_W$ equal to ${\rm sgn}\{ {\rm Det} ( \hat {\bm A} )\}$. In this case all the Weyl nodes obtained by the symmetry transformations are characterized by the same chirality. In the absence of a tilt, $d_0(\bm k) = 0$, the tensor ${\bm \gamma}$ is isotropic: the off-diagonal components are absent, while the diagonal components coincide and are equal to the contribution of a single node (\ref{Gamma00}) multiplied by the number of vectors $n$ in the star of the vector ${\bm k}_W$ if this star contains the vector $-{\bm k}_W$, and $2n$ if ${\bm k}_W$ and $-{\bm k}_W$ belong to different stars. The doubling is due to the symmetry to the time inversion which transforms the point ${\bm k}_W$ to $-{\bm k}_W$ preserving the chirality. The difference in the diagonal components allowed by the symmetry is obtained taking account of the tilt.

To calculate the off-diagonal components $\gamma_{\alpha \beta}$ in crystals of the symmetries 
C$_1$ and C$_{2}$, one needs to take into account the tilt with nonzero coefficients $a_{\alpha}$ and $a_{\beta}$. In the Hamiltonian with linear-${\bm k}$ terms, the off-diagonal components $\gamma_{xy} = -\gamma_{yx}$ in the classes C$_3$, C$_4$ and C$_6$ are not obtained, even with allowance for the tilt. As shown below, in this case it is necessary to include the higher powers of ${\bm k}$ in the Hamiltonian.

{\it The gyrotropic classes containing reflection planes}, only linear-${\bm k}$ terms in the Hamiltonian are taken into account. The off-diagonal components of $\gamma_{\alpha \beta}$ in the groups C$_s$, C$_{2v}$, S$_{4}$ and the diagonal components $\gamma_{xx}= - \gamma_{yy}$ in the groups S$_4$,  D$_{2d}$ arise in the calculation with allowance for the tilt. In the groups 
C$_{3v}$, C$_{4v}$ and C$_{6v}$, nonzero off-diagonal components do not appear in the linear Hamiltonian model.

Thus, the six gyrotropic classes C$_n$, C$_{nv}$ ($n = 3,4,6$) stand apart from the rest: for them the components $\gamma_{xy}= - \gamma_{yx}$ can be obtained by adding, to the spin-dependent part of ${\cal H}$, terms of the  second or third order in ${\bm k}$. In the simplest case, this condition is satisfied by a Hamiltonian of the mixed form
 \begin{eqnarray} \label{mixed}
&& d_x= \beta k_x + D k_y k_{\perp}^2\:,\: d_y= \beta k_y + D k_xk_{\perp}^2\:, \\ && d_z= \beta k_z  \:,\: d_0 = a_x k_x + a_y k_y\:, \nonumber
 \end{eqnarray} 
where $k_{\perp}^2 = k_x^2 + k_y^2$. In this case we obtain for the Berry curvature
\begin{eqnarray}
 \Omega_{x} &=& \frac{\beta}{2d^3}\left[ \left( \beta^2 - D^2 k_{\perp}^4 \right)k_x + 2 \beta D (k_x^2 - k_y^2) k_y \right]\:, \nonumber\\
 \Omega_{y} &=& \frac{\beta}{2d^3}\left[ \left( \beta^2 - D^2 k_{\perp}^4 \right) k_y - 2 \beta D (k_x^2 - k_y^2) k_x\right]\:, \nonumber \\
\Omega_{z} &=&  \frac{\beta k_z}{2d^3}\left( \beta^2 - 3 D^2 k_{\perp}^4 + 4 \beta D k_x k_y
\right) \:.\nonumber
 \end{eqnarray} 
 In the ${\bm k}\cdot {\bm p}$ method, the cubic terms in Eq.~(\ref{mixed}) arise from the contribution of remote bands in the third order of perturbation theory and, therefore, can be considered as being small compared to the linear terms. The components $\gamma_{xy}= - \gamma_{yx}$ become nonzero in the first order in $D$ with allowance for the tilt with $a_x^2 \neq a_y^2$.

\subsection{Circular photocurrent in crystals of the C$_{4{\lowercase{v}}}$ symmetry}

We assign number 1 to one of the Weyl nodes ${\bm k}_{W1}$ lying in the region of the Brillouin zone with the positive components ${\bm k}_{W1,\alpha} > 0$. Bearing in mind 8 elements of spatial symmetry and the time-inversion symmetry, we have 16 equivalent nodes. To analyze the net electric photocurrent, it is sufficient to consider two more nodes ${\bm k}_{W2}$ and ${\bm k}_{W3}$ obtained from the node ${\bm k}_{W1}$, respectively, by the reflection $\sigma_y$ in the plane perpendicular to the $y$ axis  and the rotation C$_4$ around the fourth-order axis. Upon passage from the node 1 to the node 2, in the effective Hamiltonian given by Eqs.~(\ref{Hamilt}), (\ref{mixed}) the coefficient $\beta$  changes its sign, the coefficient $D$ does not change, and the function $F(k_x, k_y, k_z)$ is transformed to $F(k_x, - k_y, k_z)$. When passing to the node 3, the coefficient $\beta$ is unchanged, the coefficient $D$ changes the sign, and the function $F(k_x, k_y, k_z)$ is transformed into $F(k_y, -k_x, k_z)$. In the presence of the tilt but for $D = 0$, the contributions of nodes 1 and 3 cancel each other, and no electric photocurrent is generated. An additional allowance for cubic terms is sufficient to get nonvanishing the sum over 16 nodes. Particularly, the account for this term in $\Omega_x$ leads to the current
\begin{equation} \label{jykappax}
j_y \propto \varkappa_x \frac{e^3 |{\bm E}|^2}{(\hbar \omega)^2}  \beta^2 D \sum\limits_{\bm k}
k_y ^2 (k_x^2 - k_y^2) \tilde{F}({\bm k}) \delta(2d - \hbar \omega)\:,
 \end{equation}
where $ \tilde{F}({\bm k}) = F(k_x,k_y,k_z) - F(k_y, -k_x, k_z)$. Comparable contributions come from cubic terms in the argument of the $\delta$-function and the group velocity $\partial d({\bm k})/\partial (\hbar k_{\alpha})$. According to Eq.~(\ref{jykappax}), a measurement of the circular photocurrent allows us to determine the sign of  $D$ rather than the sign of the coefficient $\beta$ which determines the chirality of the Weyl node for $D = 0$.

\section{Photocurrents in the presence of a magnetic field}

One more effect specific for gyrotropic media is a magneto-induced photocurrent independent of the light polarization. In this effect,  {\it in the linear in magnetic field approximation}, the polar vector $-$ the photocurrent density $-$ is related with the axial vector $-$ the magnetic field. For example, the following currents are generated in crystals of the C$_{2v}$ symmetry
\begin{eqnarray} \label{phenomen}
j_x &=& \left( S_{xx} B_x + S_{xy} B_y \right) |{\bm E}|^2 ,\\
j_y &=& - \left( S_{xx} B_y + S_{xy} B_x \right) |{\bm E}|^2\:, \nonumber
\end{eqnarray}
where the Cartesian coordinate system is chosen with the axis $z \parallel C_2$ and the axis $x$ composing the angle 45$^{\circ}$ with the reflection planes $\sigma_v$. 

{\it In the quantized magnetic field}, the current flows in the field direction only. Therefore, the transverse effect of the current $j_x$ generation in the field ${\bm B} \parallel y$ or the current $j_y$ in the field ${\bm B} \parallel x$ is suppressed. With account for the helicity-dependent photocurrent, we obtain the following macroscopic equations
\begin{eqnarray} \label{phenomen2}
j_x &=& \left[ S_{xx}(|B_x|) B_x + \gamma_{xx} (|B_x|) \varkappa_x \right]|{\bm E}|^2 \:,\\
j_y &=& -\left[ S_{xx}(|B_y|) B_y + \gamma_{xx} (|B_y|) \varkappa_y \right]|{\bm E}|^2 \:,\nonumber
\end{eqnarray}
where the coefficients $S_{xx},  \gamma_{xx}$ are even functions of the magnetic field strength. The detailed calculation of these coefficients and frequency dependence of the magneto-induced photocurrent will be done elsewhere. Here we briefly consider the photocurrent generated for direct optical transitions between the ground (chiral) and the first excited magnetic subbands under unpolarized excitation. In the quantizing magnetic field $\bm B \parallel y$, the energy dispersion in these subbands has the form~\cite{typeII_mcond,Tchoumakov2016} 
\begin{eqnarray}
	E_0 &=& [a_y- \text{sgn}({\cal C} B_y) \hbar\tilde{v}_0 ] k_y\:, \\
	E_1 &=& a_y k_y +  \hbar\sqrt{(\tilde{v}_0 k_y)^2 + \tilde{\omega}_c^2}\:.
\end{eqnarray}
Here $\tilde{v}_0 = v_0 / \gamma$, $v_0 = \beta/\hbar$ is the Weyl velocity in the absence of the magnetic field,
\begin{equation}
	\tilde{\omega}_c=\tilde{v}_0 \sqrt{2|eB_y|\over \hbar c },
	\quad \gamma = {1 \over \sqrt{1-(a_x/v_0)^2} },
\end{equation}
$a_x, a_y$ are the coefficients in the ${\bm k}$-linear expansion of the tilt term in Eq.~(\ref{Hamilt}) ($|a_{x,y}|<v_0$), and the coefficient $a_z$ is taken equal to zero for simplicity. We consider here one of the possible mechanisms of magneto-induced photocurrent in the Weyl semimetal of the C$_{2v}$ symmetry, where the points ${\bm k}_{W}$ lie in the plane $k_z=0$, two of them are characterized by the chirality $\cal{C}$ and the pairs of the coefficients ($a_x,a_y$), ($-a_x,-a_y$), and two others are characterized by the chirality $- \cal{C}$ and the pairs of coefficients ($a_y, a_x$), ($-a_y,- a_x$). In this mechanism, the optical transitions $E_0 \to E_1$ are allowed, depending on the sign of the field component $B_y$, predominantly in the Weyl nodes with the chirality 1 either $-1$. The photoelectrons return to the subband $E_0$ within the energy relaxation time $\tau_\varepsilon$, and only a minor part of them are scattered into a valley of the Weyl node with the opposite chirality. At zero temperature, at chemical potential $\mu>0$ lying below the bottom of the excited subband ($\mu < \hbar\tilde{\omega}_c$), and for frequencies $\omega_- < \omega < \omega_+$, where $\omega_\pm = \sqrt{(\mu/\hbar)^2+\tilde{\omega}_c^2} \pm \mu/\hbar$, the photocurrent density is given by Eq.~\eqref{phenomen2}, where $|\bm E|^2 = |E_x|^2 + |E_z|^2$, 
\begin{equation}
	S_{xx} = {\cal C} {e^3 \over 8 \pi \hbar^2} {\tau_\varepsilon \tau\over\tau_1}  [\Phi(a_x^2)-\Phi(a_y^2)]\:,
\end{equation}
the function $\Phi$ is determined as follows
\begin{equation}
\label{Phi}
	\Phi(a_x^2)= {\eta \over |B_y|} \left( {\tilde{\omega}_c\over\omega}\right)^2 \exp{\left[-\left( {a_x \omega \over \hbar v_0 \tilde{\omega}_c} \right)^2 \right]},
\end{equation}
$\eta = 1+\gamma +\gamma^2 -\gamma^3$, $\tau$ è $\tau_1$ are the times of elastic scattering between the monopoles for carriers in the ground and excited subbands, respectively ($\tau_\varepsilon \ll \tau, \tau_1$). One can see that the photocurrent remains nonzero after summation over the monopoles if $a_x^2 \neq a_y^2$.

L.E.G. and E.L.I. acknowledge the financial support of the Russian Science Foundation
(Project No. 17-12-01265).

\end{document}